\documentclass[12pt]{article}
\begin{document}
\title{Photon Spin, Zero-point Energy and Black-body Radiation}
\author{S.C. Tiwari \\
Institute of Natural Philosophy \\
1 Kusum Kutir, Mahamanapuri \\
Varanasi 221005, India }
\maketitle
\begin{center}
\textbf{{\large{}Abstract}}
\end{center}

	A critical review of the obscure nature of the contribution of 
spin energy to the energy of the electromagnetic field is presented. 
It is proposed that the total energy of a photon $h\nu$   comprises of 
kinetic and spin parts each equal to  $h\nu$/2. Classical magnetic 
field is reinterpreted as angular momentum flux of photon fluid. 
The black-body radiation law is revisited in the light of new 
significance of the zero-point energy proposed here.

\section{\bf Introduction}

Planck's law for black-body radiation spectrum presented on 19 October, 1900 before the German Physical Society
 marks the beginning of the quantum era: the defining elements of the first quantum 
theory of radiation are Planck's 
assumption that simple Hertzian oscillators of radiation possess discrete energy, the result that the average 
energy of an oscillator of frequency  $\nu$  is an integral multiple of $h\nu$  , and the assumption that 
the smallest unit of energy exchanged between matter and radiation is $h\nu$.  Planck's constant h is, similar
 to the Boltzmann constant k, a constant of nature. It was only in 1924 that a satisfactory derivation of 
Planck's law based on purely quantum nature of radiation was given by Bose. The new statistics proposed by 
Bose for photon gas was immediately applied by Einstein to the ideal gas of matter and it has come to be 
known as Bose-Einstein statistics. A nice historical review is given by Whittaker \cite{1}. Most of the modern 
textbooks of quantum theory and quantum optics briefly mention the historical significance of the Planck's law 
in the so called quantum revolution, and tend to overlook the conceptual problems which were faced by Planck 
and Einstein in reconciling the light quantum hypothesis with the classical electromagnetic field theory. 
Physical reality of light quantum (named photon by Lewis in 1926) and   its properties were uncertain during most 
of the period when the foundations of quantum theory were laid, therefore it is not only instructive to get 
acquainted with the struggle that preceded the development of wave mechanics and matrix mechanics but also 
rewarding to explore the possibility of reviving some of the old ideas for resolving contemporary issues. 
Post offers a significant  critique on the Copenhagen interpretation of quantum mechanics \cite{2} highlighting the fact 
that the presumably typical quantum result i.e. the zero-point energy was introduced first by Planck for 
an ensemble of phase randomized oscillators. Post's monograph also made me aware of the ingenious approach 
to the black-body radiation law initiated by Boyer in 1969 \cite{3}. Though the mainstream physics has 
largely ignored such efforts it would have been natural to expect that in the light of single photon 
experiments and advances in quantum optics there would be a renewed interest in such fundamental questions.
Contrary to the prevalent scenario in which quantum mysteries are supposedly serving as a ‘resource’ for 
quantum information science, a critical review \cite{4} has led me to the conclusion that even today we do 
not have a clear and unambiguous answer to the question: What is photon? In the process of trying to 
understand this question a startling fact has emerged: photon spin angular momentum has played no role in 
the   derivation of Planck's law excepting, of course, the passive polarization state counting in the 
derivation of Bose. In the present article an attempt is made to gain new insights into the black-body 
radiation physics considering the spin of photon.

It is universally recognized that Planck introduced the quantum of radiation energy, however equally 
important was his use of entropy in the problem of black-body radiation \cite{1}. It is worth mentioning 
that earlier Boltzmann proved the empirically established Stefan's law using classical radiation theory 
and second law of thermodynamics. Imprint of Planck's approach can be found in several derivations given 
by Einstein in which entropy played a crucial role; Knight and Allen \cite{5} give a succinct review of 
Einstein’s attempts beginning with his 1905 paper on photoelectric effect, and also reprint the English 
translation of his 1917 paper. In the photoelectric effect paper Einstein drew analogy between the 
entropy change for an ideal gas whose volume is varied isothermally and the similar change in entropy 
of radiation satisfying the Wien's law: this suggested that in the high frequency limit the radiation 
has corpuscular nature. The calculation of energy fluctuation for the black-body radiation using 
Planck's law in 1909 showed that large wavelength radiation had wave nature while in high frequency 
limit the radiation showed particulate aspects: this was the first enunciation of wave-particle duality 
for the radiation. The hypothesis of spontaneous and stimulated emission of radiation was put forward 
by Einstein in the important paper of 1917; it is interesting that immediately after his struggle to 
obtain the field equation for gravitation ended in 1915 he returned to the problem of light quanta. 
The existence of radiation pressure was known since the time of Maxwell, but directed momentum for photon 
and the exchange of momentum between radiation and matter during the interaction process appeared for the 
first time in this paper. It is noteworthy that Bohr's quantum theory of   atomic spectra and classical 
theory of Doppler effect both were used by Einstein in this paper.

A little known but important fact is that Planck himself was unwilling to discard classical theory of 
Maxwell, and in 1911 proposed a new hypothesis: radiation behaves as a classical wave in free space 
while emission process is discontinuous. In Planck's second theory the average energy is in excess by 
the amount of $h\nu$/2   at all temperatures to the one corresponding to the Planck's law, and at 
absolute temperature zero, the oscillator has average energy of $h\nu$/2. The  zero point energy 
associated with quantum vacuum fluctuations represents a typical counter-intuitive feature of modern 
quantum field theories, and has observational implications like Lamb shift and Casimir force. 
Whittakar, Post and Boyer in their own way have rightly emphasized its origin in the work of Planck. 
Boyer draws attention to the papers of Einstein and Hopf (1910) in which equipartition of energy in 
classical theory led to the Rayleigh-Jeans law. Following Einstein-Hopf approach, questioning the 
equipartition theorem of energy and introducing an additional hypothesis of a Lorentz-invariant spectrum 
of zero-point classical radiation Boyer \cite{3} arrives at the Planck's law. Whittaker notes that 
Einstein and Stern in 1913 derived the Planck's law using zero-point radiation.. It seems this work has 
escaped the attention of Boyer; it would be of interest to know its details.

The landmark paper of Bose (1924) had essentially three new ingredients: the  statistics assuming 
indistinguishability of photons in the photon gas, partitioning of phase space into the finite-sized 
(of size $h^3$)  discrete cells, and the use of two polarization states for photon. Note that besides these 
Bose held total energy fixed while number of photons was not conserved. The third ingredient in the Bose's 
derivation has not received adequate attention; it is unfortunate that Pais \cite{6} considers the 
derivation given by Bose as 'a successful shot in the dark', and remarks that the factor of 2 for 
polarization counting   was done with slight hesitation. Ramaseshan has placed on record reflections 
of Bose on this issue \cite{7}, and quoted from 1931 paper of Raman and Bhagawantam suggesting that 
(contrary to the views of Pais) Bose had even anticipated the spin angular momentum for photon of 
value $\pm\hbar$. Here $\hbar$   is $h/2\pi$.   Einstein appears to have been in an unsure state of 
mind on the role of angular momentum in quantum theory; in this regard Pais quotes a letter written 
by Einstein to P. Ehrenfest in 1926.  I have critically reviewed some of these aspects in Chapter 5 
of the monograph \cite{8}. I think a fairly reasonable conclusion can be drawn: unlike the key role 
of linear momentum of photon, the spin has been assigned a passive role in the derivation of Planck's 
law in spite of the  fact that black-body radiation is unpolarized.

Recently it dawned on me that both for photon and classical radiation the energy associated with spin 
or rotation has remained obscure. In \cite{4} a radical revision of the interpretation of classical 
electrodynamics has been suggested: electromagnetic field tensor itself represents the angular momentum 
tensor of photon fluid. Concerning the single photon it is argued that simple oscillator model cannot    
take into account  spin, and a new hypothesis is put forward: total energy $h\nu$   is equally divided 
into its translational energy (corresponding to momentum $h\nu$/c) and rotational energy 
(for spin $\hbar$) endowing a new significance to the zero-point energy $h\nu$/2. Some considerations of 
photon model can be found in \cite{9} and its application to understand Doppler effect is given in \cite{10}. 
In the next section the problem of energy associated with spin is elucidated, and a possible resolution 
is discussed in Sec. 3. Derivation of Planck's law is approached afresh in the light of these considerations 
in Sec. 4, and concluding remarks constitute the last section.

\section {\bf Spin and Energy}
Energy, momentum and angular momentum of the electromagnetic field and the conservation laws 
are well known, and find standard treatment in the  textbooks. Starting from the Maxwell field equations 
or using the symmetry principles in the action integral and associated Noether currents one 
arrives at these conservation laws, see e.g. Jackson's book \cite{11}. In this section certain salient 
features related with the conceptual problems are highlighted. To fix the   notation, in relativistic 
formulation Greek indices run from 0 to 3 while $x^0$ is ct, and Latin indices run from 1 to 3; 
four-volume element is $d\tau (= dx^0 dx^1 dx^2 dx^3)$, and space volume is dV. The Lagrangian density 
for the electromagnetic field is given by
\begin{equation} 
L = -(1/16\pi) F^{\mu\nu}F_{\mu\nu}
\end{equation}
Here the electromagnetic field tensor

\begin{equation}
 F_{\mu \nu} = \partial_{\mu} A_{\nu} - \partial_{\nu} A_{\mu}
\end{equation}

The action
\begin{equation}
I = \int L d\tau
\end{equation}
is a Lorentz-invariant scalar, and the infinitesimal coordinate  translation $x^\mu$ to
$x^\mu+\delta x^\mu$ leads to the covariant conservation law for the  canonical energy-momentum tensor 
$T^{\mu\nu}$. Note that we are considering source free electromagnetic field in vacuum. It is somewhat 
disconcerting to find that $T^{00}$ and $T^{0i}$ differ from the usual standard expressions for 
energy-density and momentum-density respectively given by
\begin{equation}
u=({\bf E}^2+{\bf B}^2)/{8\pi}
\end{equation}
\begin{equation}
{\bf g}=({\bf E} \times {\bf B})/{4\pi c}
\end{equation}

Integrating over all space the additional terms in $T^{00}$ and $T^{0i}$ give no contribution being 
divergence terms which are transformed to the vanishing surface integrals. Here volume integrated u 
and ${\bf g}$ transform as energy-momentum four-vector $p^\mu$.

	The canonical tensor suffers from two other formal defects: it is not symmetric, and 
is not gauge invariant. Rotational invariance,  in general, ensures angular momentum conservation. 
If we construct  angular momentum density from canonical tensor
\begin{equation}
M^{\mu\nu\lambda} = T^{\mu\nu} x^\lambda - T^{\mu\lambda} x^\nu
\end{equation}

then it is not conserved. It is possible to derive a symmetric, traceless and gauge invariant 
energy-momentum tensor. $E^{\mu\nu}$ such that
\begin{equation}
E^{\mu\nu} = T^{\mu\nu} + t^{\mu\nu}
\end{equation}

Here $t^{\mu\nu}$ is divergenceless spin energy momentum tensor that ensures angular momentum 
conservation. Interestingly $t^{\mu\nu}$  does not contribute to the total energy and momentum 
of the field  since
\begin{equation}
 \int E^{\mu 0} dV =  \int T^{\mu 0} dV
\end{equation}

Like $T^{\mu\nu}$   the spin energy-momentum tensor is also not gauge invariant, for example,
\begin{equation}
t_{\mu\nu} = \partial ^\lambda A_\mu F_{\nu\lambda}
\end{equation}
											
Let us construct angular momentum tensor from the symmetric tensor 
\begin{equation}
A_{\lambda\mu\nu} = x_\lambda E_{\mu\nu} - x_\mu E_{\lambda\nu}
\end{equation}

This differs from the angular momentum tensor $J_{\lambda\mu\nu}$ obtained from the infinitesimal 
Lorentz rotation invariance of the action I, however the distinction is unimportant since the 
difference between the two is a pure divergence term and the volume integrated angular momentum 
for both is identically equal
\begin{equation}
 \int A_{ij0}  dV =  \int J_{ij0}  dV
\end{equation} 

It would be tempting to seek division of $J_{\lambda\mu\nu}$  (or $A_{\lambda\mu\nu}$) into orbital 
and spin parts; unfortunately for a massless vector field it is not possible to identify orbital 
and spin parts in a gauge invariant manner.

	Corson in an illuminating monograph \cite{12} presents detailed discussion on the formal 
aspects of field theories and symmetries. In the following the problem of energy associated with 
angular momentum is elucidated considering the examples of plane electromagnetic wave, photon 
and multipole radiation. The vectors ${\bf E}$ and ${\bf B}$ satisfy the wave equatrion which has 
plane wave solutions. The divergence equations in the set of  Maxwell equations imply that ${\bf E}$ 
and ${\bf B}$ are both perpendicular to the direction of propagation, and the curl equations 
show that ${\bf E}$ and ${\bf B}$ are perpendicular to each other. The time-averaged energy density 
and momentum density calculated using the expressions (1) and (5) satisfy the simple relation
\begin{equation}
u = |{\bf g}|  c
\end{equation}											
It is significant that this relation holds for a monochromatic plane wave in any state of polarization. 
Beth in 1936 \cite{13} demonstrated experimentally that circularly polarized light carried angular 
momentum as suggested by Poynting; it is termed intrinsic spin angular momentum. Validity of the 
relation (12) independent of polarization seems to suggest that spin has no energy.

	In a simplified picture photon with rest mass zero is believed to satisfy the energy-momentum 
equation given by
\begin{equation}
E = |{\bf p}|  c
\end{equation}											
Note that it is analogues to Eq.(12) for plane wave. Making use of Planck's quantum hypothesis 
for energy $h\nu$   we can interpret photon momentum to be $h\nu$/c or
\begin{equation}
{\bf p} = \hbar {\bf k}
\end{equation}											
Though photon carries spin of  $\pm\hbar$  , the energy of photon once again turns out to be independent 
of the spin angular momentum. In fact, in quantum optics the polarization property of light is 
described using a polarization index (s = 1,2) in the field operators, and introducing a unit 
polarization vector basis, see \cite{14}. It can be proved that the spin angular momentum operator 
is diagonal in the basis of circular polarization in the number states of Fock space. For a plane 
light wave the spin is along the direction of propagation with the magnitude $\hbar$   times the 
difference between the number of right and left circularly polarized photons. Single photon 
state in a plane wave has momentum   $\hbar\bf k$ and spin $\hbar$ along the direction of ${\bf k}$. 
What is the energy of the spinning photon? A clear statement on this question is given by 
Kompaneyets \cite{15} on p.276: 'A quantum has one more, so to say, internal degreee of freedom, 
that of polarization. This peculiar degree of freedom corresponds to the "coordinate" $\sigma$, 
taking only two values $\sigma$ = 1 and $\sigma$ = 2. The energy does not depend upon $\sigma$'.

	Following Jackson \cite{11} we can represent the general solution of the   Maxwell field 
equations in terms of the vector spherical harmonics ${\bf L} Y_{lm}$ where  ${\bf L}$ is the orbital 
angular momentum operator and   $Y_{lm}$ is the spherical harmonics of order (l,m). Use is made of 
transverse electric and magnetic multipole fields. For the radiation field time-averaged energy 
density is calculated using Eq. (4) and the angular momentum density is calculated from the 
expression ${\bf r} \times {\bf g}$. We omit the details and note that in \cite{11} the ratio of the 
z-component of the differential of angular momentum and energy in a spherical  shell between  
r  and    r + dr  is shown to be   $m/\omega$. A quantum mechanical interpretation is indicated, 
'the radiations from a multipole of order ($l ,m$) carries off $m\hbar$ units of z-component of 
angular momentum per photon of energy   $\hbar\omega$'. Such an assertion looks quite attractive, 
however closer scrutiny reveals that the fact that photon has intrinsic spin $\hbar$   and that 
the energy  $\hbar\omega$ of a photon is due to its momentum  $h\nu$/c in the  standard picture 
make this result  intriguing. What is the contribution of $m\hbar$ units of angular momentum to 
the energy of photon?

Let us recall that in classical rotational dynamics of a point particles or a rigid body the 
angular momentum for pure rotation is defined with reference to a rotation axis, and the conservation 
laws can be expressed with reference to inertial frame of reference. Rotational energy of a rigid body 
defined in terms of moment of inertia and angular momentum is essentially the kinetic energy of linear 
motion of the constituent mass points of the body. However, intrinsic spinning motion does not have 
this kind of simple picture. The role of angular momentum in total energy of the classical electromagnetic 
field is more complicated and obscure as noted above.
\section{\bf New Approach}
It would be interesting to explore the possibility of developing a microscopic theory of 
electromagnetic fields in which one begins from the first principles using photon dynamics. 
In contrast numerous attempts since the advent of quantum theory have built photon picture 
from the quantization of the electromagnetic radiation, and face the conceptual problems of manifest 
Lorentz covariance, gauge invariance, localizability of photon, and physically meaningful photon 
wavefunction, see review in \cite{4,8,9}. Kobe in a nice review \cite{16} draws attention to a 1931 
Oppenheimer's effort to develop a photon wave equation without the electromagnetic fields. 
Kobe's own approach is the second quantization of the Schroedinger form of equation obtained from 
the Maxwell field equations. He constructs momentum operator from a velocity operator and relativistic 
mass of photon that seems strange as momentum operator has to be fundamental in canonical quantization 
not the velocity operator. In our  approach \cite{4} we make a radical departure and   argue that 
electric and magnetic fields are not fundamental, and these are macroscopic physical quantities 
describing an ensemble of large number of photons (some kind of a photon fluid). A truly microscopic 
foundation for electromagnetism would be our ultimate goal; here we present some progress in 
this direction addressing the question of angular momentum and energy discussed in the preceding section.

	Philosophical idea on which our approach is based endows physical reality to space 
(may be called aether!) and time is the cause of the manifest space and action itself \cite{17}. 
Neutrino(s), electron and photon are envisaged to be the 'spatio-temporal' objects and their 
observable attributes have geometrical and topological origin. Spatial disturbances 
(vibration and rotation) represent the internal fields of the spatio-temporal objects. 
It is assumed that internal motion is in synchronization with the translational motion: the 
internal time (or frequency) is also the periodicity of the translation such that the configuration 
returns to its initial form. This synchronization establishes a sort of what could be termed as 
relativistic rigidity in view of the relationship between spatial extension and time periodicity, 
and the absence of any rest state. Let us confine to the case of photon that is postulated to be 
a physically real object. It has been argued in \cite{4} that the electronic charge can be factored 
out from the Maxwell field equations rendering electromagnetic fields in purely geometrical units, 
for example, electric and magnetic fields have dimension of (length)$^{-2}$. The Lorentz force 
expression, however becomes
\begin{equation}
Force = e^2[{\bf E} + ({\bf v} \times {\bf B})/ c]
\end{equation}											
It is encouraging that it would be consistent with the approach in which $e^2$/c having the 
dimension of angular momentum is interpreted to give a mechanical significance to the electronic 
charge \cite{8}. To represent a single photon we consider the vector potential divided by e 
and multiplied by $\hbar$
\begin{equation}
{\bf a} = \hbar {\bf A} / e
\end{equation}											
A crucial step is to treat the three degrees of freedom associated with ${\bf a}$ to define internal 
rotation (spin) and momentum of the photon. Note that ${\bf a}$ has the  dimension of momentum. 
In the standard theory triad of unit vectors is often used to take into account helicity states 
and momentum of photon. Here two components of ${\bf a}$ determine spin and the third component   
represents the momentum of photon, for example, in the space-fixed reference frame z-axis may 
be assumed to be the direction of propagation. It appears that the natural mathematical language 
for the present approach is that of differential forms and cohomology. Post \cite{2} and 
Kiehn \cite{18} have alerted the physicists to the immense potential of de Rham cohomology 
for addressing fundamental questions in physics. It may be mentioned that gauge field theories 
and superstrings have found differential forms and fibre bundles quite fruitful. We refer to 
Kiehn's paper \cite{18} , Post's book \cite{2} and Chapter 6 in \cite{8} for physically motivated 
introduction to Cartan's exterior differential forms. In electromagnetism the vector potential 
can be written as a one-form $A_idx^i$  in three-dimension, while in space-time the four-vector 
potential gives the one-form A (= $A_\mu dx^\mu$) and the electromagnetic field tensor is a 
two-form F = dA. Here the exterior derivative d transforms a p-form to a (p+1)-form; using 
Cartan’s wedge product of p-differentials and contracting p indices of an antisymmetric tensor 
we construct a p-form. Besides d, one defines its adjoint $\delta$. In general a one-form A 
can be decomposed into three parts
\begin{equation}
A = d\alpha + \delta\beta + \gamma
\end{equation}											
The contribution from the harmonic component $\gamma$ in the integral of A contains the topological 
property. Kiehn calls such integrals as period integrals.

	We propose that the period integral of one-form a from Eq. (16) represents spin as 
a topological property of photon
\begin{equation}
 \oint a= \hbar
\end{equation}

It is important to realize that once the electronic charge is factored out and the charge is 
interpreted in terms of the fractional spin $e^2$/c the paradox discussed in the Appendix F of 
\cite{18} disappears: the apparent paradox relates with the distinction between mechanical action 
as a one-dimensional period and electromagnetic action to be a three-dimensional period. 
In the case of photon we have one-dimensional period integral Eq. (18) reminding us the 
Bohr-Sommerfeld quantization. Moreover the generalized momentum of a charged particle in the 
presence of  electromagnetic field becomes ${\bf p} - \alpha {\bf a}$ where $\alpha$ is the fine 
structure constant $e^2 /\hbar c$ that is a dimensionless number. Eq. (18) as a one-dimensional 
period integral is  consistent with the momentum integral. Post remarks on Aharonov-Bohm effect 
in his book \cite{2} and connects it with the flux quantization in a superconductor. In 
contrast, here fraction of momentum of photon is carried by the charged particle.

	The nature of the underlying manifold is a delicate issue because the conventional 
electrodynamics in four-diomensional space-time with a pair two-form F (of fields ${\bf E}$ and ${\bf B}$), 
impair two-form (of fields ${\bf D}$ and ${\bf H}$), and the current density three-form does not 
retain metric-free topological invariance as noted by Birss \cite{19}. In three-dimensional 
space it is possible to have a metric independence. The things are somewhat different in our 
approach as one-form is fundamental; however time has twin roles. There is a closed periodicity  
corresponding to internal rotation, and an open usual time coordinate corresponding to the 
external translational motion of photon. It is expected that the scalar potential would represent 
the energy, but it would be a zero-form.

	How do we represent spin and translational energy of photon? Note that translational 
periodicity gives rise to a sort of propagating harmonic wave, and there is no potential energy. 
We propose that the total energy is equally divided into spin energy and translational energy i.e.
\begin{equation}
h \nu = \hbar \omega/2 + h \nu/2
\end{equation}											
It is interesting to draw an analogy with a classical particle having momentum p, linear velocity 
v, angular momentum L and angular velocity $\omega$ then the kinetic and rotational energies 
are  pv/2 and  L$\omega$/2. If we let v=c, p=$h\nu$ /c and L =$\hbar$ we get back Eq. (19). 
It is at present not clear to me if we can obtain this result from the topological arguments.

	In this picture electromagnetic field is some kind of a photon fluid, therefore at least 
one additional scalar field is needed to describe photon number density. Eq. (18) indicates 
that magnetic field vector corresponding to ${\bf a}$ (multiplied by number density) would 
correspond to angular momentum flux, and total angular momentum would comprise of spin 
(topological) part and orbital part. One of the Maxwell field equations
\begin{equation}
\nabla.{\bf B} = 0
\end{equation}											
implies that ${\bf B} = \nabla \times {\bf A}$, and therefore, for pure gauge field ${\bf A}$ equal 
to the gradient of a scalar field, ${\bf B}$ is zero. In the unit of  $\hbar$, let us denote 
the magnetic field by ${\bf b}$, then we have
\begin{equation}
\int {\bf b}.d{\bf S} = {\bf L} + {\bf S}
\end{equation}											
in the present interpretation. What does the electric field ${\bf e}$ represent? Noting that 
the electromagnetic field tensor is a second rank  antisymmetric tensor, and the interpretation 
of ${\bf b}$ as angular momentum vector we have argued \cite{4} that the field tensor be 
interpreted as angular momentum four-tensor. The electric field would correspond to time-space 
components of the angular momentum four-tensor. In fact, the definition of ${\bf E}$
\begin{equation}
{\bf E} = - \nabla \phi - \partial {\bf A}/\partial{(ct)}
\end{equation}											
in the standard electromagnetic theory \cite{11} shows that ${\bf E}$ is a true vector 
(changes sign under space reflection or parity), and the peculiar combination of time and 
space derivatives is not relativistically covariant. Drawing the analogy with the form of the 
angular momentum four-tensor which has the components \cite{20}	
\begin{equation}
M^{0i} = c \Sigma (t{\bf p} - Energy {\bf r} /c^2)
\end{equation}											
it is proposed that ${\bf e}$ represents the motion of the photon fluid as a whole. For details 
of the conservation law of $M^{\mu\nu}$ for a closed system of particles
\begin{equation}
M^{\mu\nu} = \Sigma ( x^\mu p^\nu - x^\nu p^\mu)
\end{equation}
where summation is over all particles of the system, we refer to \cite{20}
											
Unlike the interpretation of magnetic field as angular momentum flux of the photon fluid, 
the electric field vector does not seem to have straightforward physical significance. If 
${\bf e}$ represents the translation of photon fluid then similar to Eq. (19), for a single photon 
Eq. (4) could be interpreted as a sum of translational and rotational energy density of the photon fluid. 
Note the form of the expression of energy for a classical particle
\begin{equation}
Energy = p^2/2m +L^2/2I
\end{equation}											
On the other hand, assuming the definition (22) and noting the interpretation that vector 
potential and scalar potential represent momentum and energy density respectively the electric 
field would represent force density.
\section{\bf Black-Body Radiation}
Planck's second theory gave statistical origin of the zero-point energy, and in modern theories 
it is a typical quantum effect. Eq. (19) offers a third origin: $h\nu$/2  energy corresponds to 
the  $h\nu$ /c linear momentum  and $\hbar\omega$/2  energy to the spin angular momentum of $\hbar$
for a single photon. In any process that depends only on the momentum exchange  the energy for 
a photon has to be taken  $h\nu$ /2 not $h\nu$ as is done conventionally. Though black-body radiation 
law is well established, we revisit it in the light of our hypothesis. It is significant that 
Boyer \cite{21} has given four derivations of the radiation law, and in each of them zero-point 
radiation plays a crucial role. In his approach the zero-point radiation is a random classical 
radiation. If 'quantum ideas' means discreteness then our approach belongs to extreme quantum 
domain since space itself is discretized; however it has nothing to do with the standard quantum theory.
 
	In all derivations momentum exchange is the key process for attaining thermodynamical 
equilibrium. Whether one is considering classical radiation or one is considering light quantum, 
factor of 2 in essence arises due to the polarization states. Bose calculates phase space volume 
and multiplies it by 2 to account for two polarization states of photon. Let us consider Eq. (19) 
then the first expected change would be that for momentum  $h\nu$/c for a photon we have to 
take energy $h\nu$/2. Retaining other assumptions made by Bose the final result for the energy 
density $u(\nu) d\nu$  within the frequency interval $d\nu$ is given by 
\begin{equation}
u(\nu)d\nu = (h\nu/2)\frac {8\pi\nu^2 d\nu}{c^3(e^{h\nu/2kT} - 1)}
\end{equation}											
It is easy to show that for small $\nu$ this expression goes over to the Rayleigh-Jeans law. 
Integrating over all frequencies it can be verified that the total energy density satisfies 
the Stefan-Boltzmann law. However there is something amiss here as Eq. (26) does not agree with 
the Planck's formula.

	An important characteristic of black-body radiation is that it is unpolarized. 
Recently the nature of unpolarized light has been the subject of renewed significance, see Lehner 
et al \cite{22} for a critical discussion and review. Assuming spin or polarization correlated 
photon pairs such that each single object (i.e. the pair) has spin zero, the ensemble of these 
pairs would certainly be unpolarized. If in the thermodynamical equilibrium of the black-body 
radiation photon pairs are assumed to be the basic entities in the momentum exchange process 
then the corresponding energy becomes h$\nu$, and the multiplication by 2 gets added significance 
counting two photons with opposite spin in the pair. Obviously the Planck's formula is 
recovered with a new interpretation. Mandel and Wolf \cite{14} discuss polarization using 
the coherence matrix, and show that for each propagation vector the black-body radiation is 
unpolarized. Pairs of photons with given momentum as proposed by us are also unpolarized. 
Taking note of the characteristic polarization entangled states for which each one of the photons 
is unpolarized \cite{14}, it is possible that in the black-body radiation the photon pairs at 
the source are generated in polarization correlated states.

	Note that spin energy has not entered into the discussion of the Planck's law: is 
there no role of this dormant energy? A careful reflection shows that at absolute temperature zero, 
the zero-point radiation spectrum has a plausible origin in terms of the spin energy  $\hbar\omega$/2         
of photons with momentum in a sort of frozen state resembling with that of Bose-Einstein condensate 
which is a zero momentum state. Besides this we envisage the possibility of  imparting rotation 
to small particles suspended in the  black-body radiation enclosure in which polarized light beams
are injected from outside source. Physically one would expect spin energy exchange to contribute 
in the thermodynamical process in this modified black-body enclosure of mixed i.e. unpolarized+polarized
radiation.
\section{\bf Conclusion}
	I have been investigating the well known foundational problems of electrodynamics for past more 
than two decades which have defied satisfactory solution in spite of the great efforts put by e
minent physicists; the present paper is a contribution in this continuing process of my understanding. 
The main results of the present work are the definitive interpretation of magnetic field as angular 
momentum flux, the hypothesis that the total energy of a single photon is equally divided into kinetic 
and spin parts, and the new insight gained from the black-body radiation formula using a 
radically different interpretation of zero-point energy.

	Some of the arguments are still tentative, and the inconclusive interpretation of 
the electric field is unsatisfactory. In a complete theory it is expected that electric field 
and the Maxwell equations would emerge from photon fluid dynamics representing the property of the 
fluid. Two options to tackle this problem are being explored: (1) assuming the analogy with the 
rotating superfluid and treating photons as vortices, the microscopic theory of superfluidity could 
be applied to the photon fluid case, and (2) since photon is proposed to be a space-time structure, 
the abstract space-time flow of null lines in a geometrical framework offers another approach. 
In the curved space-time geometry there is an important result: null lines are subjected to 
vorticity and shear. The flow of null vector lines gets separation and expansion as a geometrical 
effect. Since physically photon trajectory is a null line, and Maxwell field equations satisfy 
general covariance, geometry and topology of space-time could throw some light on developing 
the photon fluid paradigm for electromagnetism.

\section{ACKNOWLEDGEMENT}

	The library facility at the Banaras Hindu University is acknowledged.

\end{document}